\documentstyle[twocolumn,prl,aps]{revtex}
\begin{document}
\draft
\title{Where does the Rho Go?  Chirally Symmetric Vector Mesons in the
Quark--Gluon Plasma.}
\author{Robert D. Pisarski}
\address{
Department of Physics, Brookhaven National Laboratory,
P.O. Box 5000, Upton, New York 11973-5000, USA}
\date{\today}
\maketitle
\begin{abstract}
If the phase transition of $QCD$ at nonzero temperature is dominated by
the (approximate) restoration of chiral symmetry,
then the transition might be characterized using
a gauged linear sigma model.  Assuming that vector
meson dominance holds, such
sigma models predict that at the temperature of chiral restoration, the
pole mass of the thermal $\rho$
meson is greater than that at zero temperature;
in the chiral limit and in weak coupling
this mass is $\sim 962 \, MeV$.  The width of the
thermal $\rho-a_1$ peak is estimated to be about
$200 - 250 \, MeV$.
\end{abstract}
\pacs{BNL Preprint BNL/RP-951, January, 1995. }
\begin{narrowtext}

When the quark--gluon plasma is produced
by the collision of large nuclei at ultrarelativistic energies, such as at
$RHIC$ and $LHC$, the crucial question is how to detect its
presence as the plasma expands and cools into ordinary hadronic matter.
A promising signal is to look at the production of dileptons, since
they escape from the fireball essentially without interaction.
The most prominent feature of the dilepton spectra are the peaks from
their coupling to vector mesons.

Vector mesons can be classified into two types.  For mesons such as
the $\rho$, their lifetime is so short that they decay within the plasma.
Consequently, the shift in their mass and width from interactions
in the plasma --- the nature of the ``thermal $\rho$'' ---
are in principle observable \cite{rdp1}.  The second type of mesons are
those whose lifetime is so long that they decay outside of the plasma,
like the $J/\Psi$.  Then any
shift in the mass or width is not observable,
but one can measure a relative depletion in the height
of the peak \cite{ms}.

In this Letter I investigate the nature of the thermal $\rho$ within the
context of a gauged linear sigma model \cite{old}.  Several other authors have
conducted similar studies in these \cite{vect},\cite{lsy} and other
\cite{shuryak}-$\!\!$\cite{eib} models.
The principal point herein is that at least in weak coupling, a general
feature of gauged linear sigma
models is that at the point where chiral symmetry is
restored, the mass of the thermal $\rho$ is {\it greater}
than that at zero temperature.
The shift in the $\rho$ mass can be relatively large, on the order of
$T_\chi$, where $T_\chi$ is the temperature for the restoration of chiral
symmetry.

I work with two flavors, assuming that the effects of the
axial anomaly are always large,
so the global chiral symmetry is $SU(2)_l \times SU(2)_r$.
Introducing the matrices $t^0 = {\bf 1}/2$ and
$t^a$, $tr(t^a t^b) = \delta^{a b}/2$, the scalar field $\Phi$ is
$$
\Phi = \sigma \, t^0 + i \vec{\pi} \! \cdot \! \vec{t} \; ;
$$
$\vec{\pi}$ is the $J^P=0^-$ isotriplet pion field and
$\sigma$ a $0^+$ isosinglet field.
For the left and right handed vector fields I take
$$
A^\mu _{l,r} = (\omega^\mu \pm f_1^\mu)t^0
+ (\vec{\rho}^{\, \mu} \pm \vec{a}_1^{\, \mu})\!\cdot\! \vec{t}
$$
where $\omega$ and $\vec{\rho}$ are $1^-$ fields,
and $f_1$ and $\vec{a}_1$ are $1^+$ fields.
According to the principle of vector meson
dominance \cite{old}, the
dimensionless couplings of the vector fields to themselves
and to $\Phi$ are exclusively those which
follow by promoting the global chiral symmetry to a local symmetry.
Introducing the coupling constant $g$ for vector meson dominance,
the appropriate covariant derivative and field strengths are
$D^\mu \Phi = \partial^\mu \Phi - i g (A^\mu_l \Phi - \Phi A^\mu_r)$
and $F_{l,r}^{\mu \nu}
= \partial^\mu A_{l,r}^\nu - \partial^\nu A_{l,r}^\mu
- i g [A_{l,r}^\mu , A_{l,r}^\nu] $.  The effective lagrangian is then
$$
{\cal L} = tr \left( \left|D^\mu \Phi\right|^2
- \mu^2 |\Phi|^2 + \lambda (|\Phi|^2)^2 - 2 h t^0 \Phi \right. \;
$$
\begin{equation}
\left. \; + \frac{1}{2} (F_l^{\mu \nu})^2 + \frac{1}{2}(F_r^{\mu \nu})^2
+ m^2 \left( (A_l^\mu)^2 + (A_r^\mu)^2 \right) \right) \; .
\label{e1}
\end{equation}
Including $g$, the parameters of the model are a mass
squared $- \mu^2$, which drives spontaneous symmetry breaking
at zero temperature, a dimensionless scalar coupling $\lambda$,
a background field $h$ to make the pions massive, and a mass term
$\sim m^2$ for the gauge fields \cite{singlet}.
Much of the physics of this lagrangian can be understood from the kinetic
term for the scalar field,
$$
tr \left( |D^\mu \Phi|^2 \right) =
\frac{1}{2} \left( \left(\partial^\mu \sigma + g \vec{a}_1^{\, \mu}
\! \cdot \! \vec{\pi}\right)^2 \right.
$$
\begin{equation}
+ \left. \left( \partial^\mu \vec{\pi} + g \vec{\rho}^{\, \mu}
\!\times\! \vec{\pi}
- g \vec{a}_1^{\, \mu} \sigma \right)^2
+ g^2 \left( \sigma^2 + \vec{\pi}^2 \right) (f_1^\mu)^2 \right)
\label{e2}
\end{equation}
Because it couples to the (isosinglet) current for fermion number,
the $\omega^\mu$ field drops completely out of (\ref{e2}).  There
are interactions of $\omega^\mu$ due to effects of the anomaly,
but these are neglected in this work.

I stress how remarkable the principle of vector meson dominance
is.  If one constructs the most general
lagragian consonant with the {\it global} chiral symmetry
of $SU(2)_l \times SU(2)_r$, then instead of a one
coupling constant $g$, many more dimensionless coupling constants
are required \cite{largeN}.  Vector meson dominance limits
the breaking of the local chiral symmetry solely to soft
mass terms \cite{singlet}, such as that $\sim m^2$ in (\ref{e1}).
As I discuss at the end of this Letter,
if the principle of vector meson dominance is abandoned, then very
different predictions follow.

Of course the price paid is
that the theory is not perturbatively renormalizable.  For
a vector field with mass $m$,
in momentum space the propagator is
$\Delta^{\mu \nu}(P) = (\delta^{\mu \nu}
- P^\mu P^\nu/m^2)/(P^2 + m^2)$,
which is $\sim 1$ and so
badly behaved at large $P$.  In the present analysis this
lack of renormalizability is inconsequential.  This is because
I assume that I am always in a regime where the
temperature $T \leq T_\chi \ll m$,
and for such low temperatures the effects of
quantum vector fields should be temperature independent.

When spontaneous symmetry breaking occurs, so
$\sigma \rightarrow \sigma_0 + \sigma$,
the vector meson masses are \cite{singlet}
\begin{equation}
m_\rho^2 = m_\omega^2 = m^2 , \;\;
m_{a_1}^2 = m_{f_1}^2 = m^2 + (g \sigma_0)^2 ,
\label{e3}
\end{equation}
Further, from (\ref{e2}) $\sigma_0 \neq 0$ generates a mixing between
the $\vec{a}_1^\mu$ field with $\partial^\mu \vec{\pi}$.  This
produces a type of ``partial'' Higgs effect, whereby the standard
results in a linear sigma model are modified by ratios of
$m_{a_1}/m_\rho$:
\begin{equation}
f_\pi = \frac{ m_{\rho} }{m_{a_1}} \sigma_0 , \;\;
m^2_\pi = \frac{ m^2_{a_1} }{m^2_\rho} \frac{h}{\sigma_0} , \;\;
m^2_\sigma = \frac{h}{\sigma_0} + 2 \lambda \sigma_0^2 .
\label{e4}
\end{equation}
In $MeV$ I use the values $f_\pi = 93$, $m_\pi = 137$,
$m_\rho = 770$, and $m_{a_1} = 1260$.
Notice that the value of ratio $m_{a_1}/m_\rho \sim 1.6$ is significantly
larger than one.  These values determine
$\sigma_0 = 152 \, MeV$, $g=6.55$, $h = (102 \, MeV)^3$,
and $m = 770 \, MeV$.
The values of the remaining parameters depend upon the value
of $m_\sigma$.  I choose two representative values \cite{sigma};
$m_\sigma= 600 \, MeV$ gives $\lambda = 7.62$ and $\mu = 412 \, MeV$,
while
$m_\sigma= 1000 \, MeV$ gives $\lambda = 21.4$ and $\mu = 700 \, MeV$.
With these values of $\lambda$ and $g$ the theory is
manifestly in a strong coupling regime.  Nevertheless, to gain a
qualitative understanding of the physics I work to lowest
order in a loop expansion.

In weak coupling it is easy
to compute the thermal masses at the temperature of chiral
symmetry restoration, $T_\chi$.  For simplicity
I work in the chiral limit, $h=0$, where
$T_\chi^2 = 2 \sigma_0^2$,
so $T_\chi = 215 \, MeV$ \cite{temp},\cite{lattice}.
At $T_\chi$ I can compute in the
symmetric phase, working from above.
A technical but crucial point is that it is necessary to compute
the self energies not at zero momentum, but on the relevant mass
shell, since this is what determines the coupling to dileptons.
Consequently, instead of the low momentum limit of the self energies,
one is interested in their limit for large momentum $P \gg T$.
Calculation shows that the $\rho$ and $a_1$ self energies
are each
$\Pi^{\mu \nu} = (\delta^{\mu \nu} - P^\mu P^\nu/P^2) (g^2 T^2/6)$,
while the $f_1$ self energy is
$\Pi^{\mu \nu} = \delta^{\mu \nu} (g^2 T^2/3)$ at large $P \gg T$.
Using $T_\chi^2 = 2 \sigma_0^2$ and (\ref{e3}),
in weak coupling at the critical temperature
the pole masses in the vector meson propagators are given by
$$
m^2_\rho(T_\chi) = m^2_{a_1}(T_\chi) =
\frac{1}{3} \left( 2 m^2_\rho + m^2_{a_1} \right)
= (962 \, MeV)^2 \; ,
$$
\begin{equation}
m^2_{f_1}(T_\chi) =
\frac{1}{3} \left( m^2_\rho + 2 m^2_{a_1} \right)
= (1120 \, MeV)^2 .
\label{e5}
\end{equation}
On the right hand side of
(\ref{e5}) and henceforth, whenever I write
a mass such as $m_\rho$ or $m_{a_1}$,
implicitly I am referring to their values at zero temperature;
any thermal pole mass is denoted by $m_\rho(T)$, {\it etc.}

Since in (\ref{e2}) the $\omega$ field does not interact with
the scalar fields,
the $\omega$ mass does not move,
$m^2_\omega(T) = m_\omega^2$; $m_\omega^2(T)$ only shifts from effects
of the anomaly.
At the very least, it is apparent that
the near degeneracy between the zero temperature masses of the
$\omega$ and the $\rho$, and the $a_1$ and the $f_1$,
is badly broken at nonzero temperature.

The width of the $\rho$ can be computed by standard means \cite{weldon};
at one loop order the only available mode is
$\rho \rightarrow \pi \pi$.  For a $\rho$ decaying at rest,
\begin{equation}
\Gamma_\rho^\chi = \frac{g^2}{48 \pi} \left( 1 + 2 n(m^\chi_\rho/2) \right)
\frac{((m^\chi_\rho)^2 - 4 (m_\pi^\chi)^2)^{3/2}}{(m^\chi_\rho)^2} \; .
\label{e6}
\end{equation}
Here $m^\chi_\pi = m_\pi(T_\chi)$ and $m^\chi_\rho = m_\rho(T_\chi)$ are
the thermal pole masses at $T=T_\chi$, and $\Gamma_\rho^\chi =
\Gamma_\rho(T_\chi)$.
This is just the standard formula for the decay width of the $\rho$,
except that there is a factor involving the Bose-Einstein distribution
function, $n(E)= 1/(exp(E/T) - 1)$, from stimulated pion emission
in a thermal bath.
At zero temperature, (\ref{e6}) gives a decay width that is about $20 \%$
too large,
$\Gamma_\rho(0) \sim 179 \, MeV$ instead of the experimental value
of $150 \, MeV$.

To obtain a somewhat realistic estimate of the
width of the thermal $\rho$, the nonzero mass of the pion must
be included.
The full problem with $h \neq 0$ and $T \neq 0$
is rather complicated, since $m_\pi^\chi \sim T_\chi$.
I adopt an approximate solution: the thermal
effects are computed in the high temperature limit, including only
the terms
$\delta{\cal L} = (\lambda T^2/2) tr(|\Phi|^2) + (g^2 T^2/12)
((\vec{\rho}^{\, \mu})^2 + (\vec{a}_1^{\, \mu})^2)$.
When $h \neq 0$
the definition of $T_\chi$ is ambiguous; I define
$T_\chi$ as the point where $m_\sigma(T)$ has a minimum with respect to
$T$.  Doing so,
for $m_\sigma = 600 \, MeV$ I find
$T_\chi = 226 \, MeV$; at $T=T_\chi$, $f_\pi^\chi = 32 \, MeV$,
$m_\rho^\chi =
978 \, MeV$, $m_{a_1}^\chi = 1002 \, MeV$, $m_\pi^\chi = 185 \, MeV$,
$m_\sigma^\chi = 221 \, MeV$, and $\Gamma_\rho^\chi = 278 \, MeV$.
For $m_\sigma = 1000 \, MeV$ I find:
$T_\chi = 221 \, MeV$; at $T=T_\chi$, $f_\pi^\chi = 23 \, MeV$,
$m_\rho^\chi = 971 \, MeV$, $m_{a_1}^\chi = 983 \, MeV$,
$m_\pi^\chi = 217 \, MeV$,
$m_\sigma^\chi = 263 \, MeV$, and $\Gamma_\rho^\chi = 248 \, MeV$.
If I assume that the $\rho$ width is too high by the same amount
at $T_\chi$ as at $T=0$, and so should be corrected by a factor of
$150/179$, I obtain $\Gamma_\rho^\chi= 233 \, MeV$ for $m_\sigma
= 600 \, MeV$
and $\Gamma_\rho^\chi= 208 \, MeV$ for $m_\sigma = 1000 \, MeV$.

The form in which I have written (\ref{e5}) is a bit
misleading, in that at leading order in weak coupling I can
eliminate $g$ entirely, to
write expressions for the masses at $T_\chi$ solely in terms
of the zero temperature masses.
Nevertheless, it should be emphasized that this is a trick only of
results to lowest order;
the corrections to (\ref{e5}) and (\ref{e6}) are a power series in
$g^2$ and $\lambda$, and so large.  Thus the above numerical values
are {\it not} meant to be taken
as predictions, but {\it only} as suggestions of the magnitude of the
possible effect.
Perhaps, however, the {\it qualitative} features of a weak coupling
analysis are reasonable.  At zero temperature the splitting
between the $\rho$ and $a_1$ masses are driven entirely
by spontaneous
symmetry breaking; it is sensible that the thermal fluctuations
which restore the symmetry are of the same order as the shift
upward in the (thermal) $\rho$ mass.  Similarly, while thermal
broadening can be very significant if the $\rho$ mass decreases,
if the $\rho$ mass increases these effects are
naturally small, since then the $\pi$'s are energetic, with
momenta significantly larger than the temperature.  One effect which
I have neglected which increases $\Gamma_\rho^\chi$ is the
thermal width of the $\pi$'s; however, a more realistic value of
$T_\chi$ is probably lower than the above
\cite{temp}, which lowers $\Gamma_\rho^\chi$.

It is also of interest to compute the shift in the pole masses
at low temperature.  In the chiral limit
we can make comparison with a general
analysis of Eletsky and Ioffe \cite{ei}, who show that the
shift in the pole masses vanishes to order $\sim T^2$ about
$T=0$.   In gauged sigma models
this holds for both the $\rho$ and $a_1$ masses \cite{lsy}.  The first
nonleading terms in the pole masses for the transverse fields are,
in the chiral limit,
$$
m^2_\rho(T) \sim m^2_\rho - \frac{g^2 \pi^2 T^4}{45 m^2_\rho}
\left(\frac{4 m_{a_1}^2 (3 m_\rho^2 + 4 p^2)}
{(m_{a_1}^2 - m_\rho^2)^2} - 3 \right) + \ldots \; ,
$$
$$
m^2_{a_1}(T) \sim m^2_{a_1} + \frac{g^2 \pi^2 T^4}{45 m_\rho^2} \left(
\frac{4 m_{a_1}^2 (3 m_{a_1}^2 + 4 p^2)}{(m_{a_1}^2 - m_\rho^2)^2} \right.
$$
\begin{equation}
\left. + \frac{2 m_\rho^4}{m_{a_1}^2 (m_{a_1}^2 - m_\sigma^2)}
- \frac{m_{a_1}^2}{m_\rho^2} \right) + \ldots \; ,
\label{e7}
\end{equation}
where $p^2$ is the spatial momentum squared of the field.
That is, while by the time of the chiral transition
the thermal $\rho$ mass goes up, and the $a_1$ mass down,
about zero temperature they {\it start} out in the opposite direction:
the $\rho$ mass goes down, and the $a_1$ up!

Putting in the values of $m_\rho$, $m_{a_1}$ and $g$,
at zero momentum, $p=0$, I find
that $(m_\rho^2(T) - m_\rho^2)/m_\rho^2 = - (2.98 \, T/m_\rho)^4$, while
$(m_{a_1}^2(T) - m_{a_1}^2)/m_{a_1}^2 = + (3.16 \, T/m_\rho)^4$
when $m_\sigma = 600 \, MeV$, and
$(m_{a_1}^2(T) - m_{a_1}^2)/m_{a_1}^2 = + (3.17 \, T/m_\rho)^4$
for $m_\sigma = 1000 \, MeV$.  These values are interesting because
the coefficients of $T/m_\rho$ on the right hand side are relatively
large:
if we push them well beyond their range of validity, to $T \sim 200 \, MeV$,
they suggest that the shifts in the thermal
$\rho$ and $a_1$ masses can be significant, on the order of $T_\chi$,
as found in (\ref{e5}).

The shift in the thermal masses at low temperature can
also be computed away from the chiral limit.
When $m_\pi \neq 0$ I find that the $\rho$
mass does not shift to $\sim T^2$, but the $a_1$ mass
does,
\begin{equation}
m^2_{a_1}(T) \sim  m^2_{a_1}
+ \frac{ g^2 m^2_\pi T^2}{4 m^2_\sigma } + \ldots \; .
\label{e8}
\end{equation}
As for the $\sim T^4$ term in the chiral limit, (\ref{e7}), when
$m_\pi \neq 0$ the $a_1$ mass starts out by going {\it up} at
low temperature.  In $QCD$,
except at the very lowest temperatures,
this correction is small relative
to that in (\ref{e7}):
$(m_{a_1}^2(T) - m_{a_1}^2)/m_{a_1}^2 = + (.46 \, T/m_\rho)^2$
for $m_\sigma = 600 \, MeV$, and
$(m_{a_1}^2(T) - m_{a_1}^2)/m_{a_1}^2 = + (.27 \, T/m_\rho)^2$
for $m_\sigma = 1000 \, MeV$.

I conclude by discussing the relationship with other approaches.
By using a gauged linear
sigma model for $T \leq T_\chi$,
implicitly I am assuming that the
behavior of $QCD$ at nonzero temperature is dominated by the
restoration of chiral symmetry, and not by deconfinement.
This accords with current
numerical simulations of lattice gauge theory \cite{lattice},
which indicates while there is no true phase transition in $QCD$,
it lies close to a chiral critical point \cite{rdp2}.

In contrast, if the phase transition were dominated by
deconfinement, then as argued initially in
\cite{rdp1}, it is conceivable that the thermal $\rho$ mass
decreases with increasing temperature.
For example, sum rule
analyses of the phase transition can be construed as
dominated by deconfinement.  Generally, such analyses find
that the thermal $\rho$ mass goes down as $T$ goes up
\cite{sumdown} (see, however, \cite{sumup});
about zero temperature, ref. \cite{eib} find that
both the $\rho$ and $a_1$ masses decrease to $\sim T^4$,
contrary to (\ref{e7}).  Using the experimental phase shifts,
Shuryak and Thorsson \cite{shuryak}
also find that the $\rho$ mass decreases,
by a small amount, at $T \sim T_\chi$.

Following Georgi,
Brown and Rho, and others \cite{brown}, have analyzed a sigma
model where the $\rho$ mass decreases monotonically with
temperature.  While their analysis uses a
nonlinear sigma model, it can be reexpressed in
terms of a linear sigma model.
Assume that the explicit mass
term for the gauge fields $\sim m^2$ in (\ref{e1}) vanishes, and that
instead the local chiral symmetry is broken {\it only}
by the term such as ${\cal L}_\kappa = \kappa \, tr(|\Phi|^2)
tr ( (A_l^\mu)^2 + (A_r^\mu)^2 )$, where $\kappa$ is a dimensionless
coupling constant.  With such a term, up to
$T_\chi$ the $\rho$ mass {\it does} decrease uniformly
with temperature; an easy calculation shows that in the chiral limit,
$m^2_\rho(T_\chi) = m^2_{a_1}(T_\chi) = (2/3) m^2_\rho = (629 \, MeV)^2$.
However, setting $m=0$
and including ${\cal L}_\kappa$ manifestly violates
the assumption of vector meson dominance, since then the local
chiral symmetry is broken by a term with a dimensionless, instead of
a dimensional, coupling constant.

In other words, which way the thermal $\rho$ goes depends crucially
upon whether or not vector meson dominance applies at nonzero temperature.
If vector meson dominance holds, the thermal $\rho$ mass goes up
by $T_\chi$; without
vector meson dominance, there is no unique prediction.

To understand the relationship to the theory at $T \geq T_\chi$,
it is necessary to remember that a constant feature of the
lattice results \cite{lattice}
is that independent of the order of the
phase transition, uniformly
there appears to be a large increase in the entropy in a narrow
region of temperature.  Such a large increase in entropy
cannot be described
by the kind of gauged linear sigma models which I have been using.
Consequently, I presume that such sigma models are valid
{\it only} to a temperature just below $T_\chi$, but {\it not} above.

In heavy ion collisions at ultrarelativistic energies,
then, if a mixed phase lives for a long
time and dominates total dilepton production, a two state signal
should appear in dilepton production.
{}From the quark-gluon phase at $T = T_\chi^+$, dilepton production
is dominated by the quark quasiparticles \cite{rdp3}, presumably
concentrated in a region below the zero temperature
$\rho$ peak.  The hadronic phase at $T= T_\chi^-$ generates a
thermal $\rho$ peak; the position of this peak is model dependent,
lying either above or below the zero temperature
$\rho$ peak, depending upon whether the assumptions of
\cite{vect}, \cite{lsy}, and this work,
or those of \cite{shuryak}, \cite{brown}, \cite{sumdown}, and
\cite{eib}, apply.

Whichever scenario applies, theoretically there are numerous
indications that if it is possible to resolve relatively
wide structure in dilepton
production in ultrarelativistic heavy ion collisions ---
on the order of $T_\chi \sim 200 \, MeV$ --- then it
might well reveal novel structure.
While experimentally this is an {\it extremely} difficult
task, the possible rewards appear well worth the effort.

I happily (if belatedly)
acknowledge that an inspirational colloquium on the quark-gluon plasma
by W. J. Willis at Yale University in 1981 originally \cite{rdp1} stimulated
my interest in this problem.
During the present investigation I benefited from discussions with
J. Bijnens, V. Eletsky, T. Hatsuda, S.-H. Lee, M. Rho, E. Shuryak,
A. Sirlin, C. Song, L. Trueman, A. Weldon, and especially S. Gavin.
This work is supported by a DOE grant at
Brookhaven National Laboratory, DE-AC02-76CH00016.

\end{narrowtext}

\begin{references}
\bibitem{largeN}
I note that vector meson dominance
is consistent with the limit of a large number of colors,
$N_c \rightarrow \infty$.
Vector meson dominance
implies $j^\mu_{em} \sim (m^2/g) \rho^\mu$; assuming that
$m \sim 1$ and $g \sim 1/\sqrt{N_c}$,
$\langle j^\mu_{em} j^\nu_{em} \rangle \sim N_c$,
as follows from the underlying quark diagrams.
Consider corrections in $1/N_c$: the $\rho$ peak, which is
of order $\sim N_c$ in height, acquires a width $\sim 1/N_c$.
At this order
$\pi^+ \pi^-$ production also contributes;
since hadrons are free at large $N_c$, this is calculable,
of order $\sim 1$.
Thus the area under both the $\rho$ peak and the
$\pi^+ \pi^-$ continuum are of order $\sim 1$.
\bibitem{temp}
Because of the factor
of $m_{a_1}/m_\rho$ which enters into $f_\pi$ in
(\ref{e4}), this is significantly
higher than the estimate of $T_\chi$ without gauge fields,
which is just $T_\chi = \sqrt{2} f_\pi = 131 \, MeV$;
it is also significantly higher than the estimate from
lattice simulations, which are $T_\chi \sim 150 \, MeV$ \cite{lattice}.
I presume this is a shortcoming of the gauged linear sigma model.
\end{references}
\end{document}